\documentclass[prc,onecolumn,aps,amsmath,amssymb,showpacs]{revtex4}

\usepackage{graphicx}
\usepackage{dcolumn}
\usepackage{bm}

\def\thalf{{\textstyle{\frac{1}{2}}}}
\def\tquar{{\textstyle{\frac{1}{4}}}}
\def\threequ{{\textstyle{\frac{3}{4}}}}
\def\fiveth{{\textstyle{\frac{5}{3}}}}
\def\twoth{{\textstyle{\frac{2}{3}}}}
\def\oneth{{\textstyle{\frac{1}{3}}}}

\begin{document}
\preprint{NUC-MINN-01/19-T}
\title{Role of Fluctuations in the Linear Sigma Model with
Quarks}
\author{\'A. M\'ocsy$^{1,2,3}$, I.N. Mishustin$^{1,2,3,4}$ and
P.J. Ellis$^3$} \affiliation{$^1$Institut f\"ur Theoretische
Physik, J.W. Goethe Universit\"at, D-60054 Frankfurt am Main,
Germany\\ $^2$The Niels Bohr Institute, Blegdamsvej 17, DK--2100
Copenhagen \O, Denmark\\ $^3$School of Physics
and Astronomy, University of Minnesota, Minneapolis, MN 55455,
USA\\$^4$The Kurchatov Institute, Russian
Research Center, Moscow RU-123182, Russia \\ }

\begin{abstract}
We study the thermodynamics of the linear sigma model with constituent 
quarks beyond the mean-field approximation. By integrating out the 
quark degrees of freedom we derive an effective action for the meson 
fields which is then linearized around the ground state including 
field fluctuations. We propose a new method for performing exact 
averaging of complicated functions over the meson field fluctuations. 
Both thermal and zero-point fluctuations are considered. The chiral
condensate and the effective meson masses are determined
self-consistently in a rigorous thermodynamic framework. At zero chemical
potential the model predicts a chiral crossover transition which 
separates two distinct regimes: heavy quarks and light pions at low 
temperatures, but light quarks and heavy mesons at high temperatures.
The crossover becomes a first order phase transition if the vacuum
pion mass is reduced from its physical value.
\end{abstract}
\maketitle

\section{Introduction}

It is commonly accepted that Quantum Chromo-Dynamics (QCD) is the
true theory of strong interactions. Therefore, in principle, it
should describe all phases of strongly interacting matter at all
densities and temperatures. In practice QCD can be exactly solved
only in some limiting cases: firstly, at very high densities and
temperatures when the property of asymptotic freedom allows a
perturbative expansion to be used, and secondly, at zero density
and high enough temperatures when it can be discretized on an
Euclidean lattice. Although some improvements at finite chemical
potential have been achieved recently \cite{Fodor:2001pe}, the
direct application of QCD at high baryon density, or for real-time
processes, is still quite problematic at present. This makes it
necessary to build effective models, which respect only general
properties of QCD, such as chiral symmetry, and operate with
effective degrees of freedom, such as mesons and constituent
quarks.

The linear sigma model (L$\sigma$M) \cite{GellMann&Levy} is one of
the most popular models of this kind which has been studied
already for several decades (see e.g. Refs.
\cite{Baym&Grinstein,BK}). It incorporates correctly the
phenomenology of low-energy strong interactions, including chiral
symmetry. At low density and temperature the matter is assumed to
be in a phase where chiral symmetry is spontaneously broken. A
phase transition that restores chiral symmetry was predicted at
high temperatures \cite{Kirzhnits&Linde} or baryon densities
\cite{Lee&Wick}. According to Ref. \cite{PW} QCD with two massless
flavors belongs to the same universality class as the $O(4)$
L$\sigma$M and therefore the phase transition is of the same order. 
In the case of nonzero quark masses chiral symmetry is
explicitly broken and universality arguments cannot be applied.
Then the character of the chiral transition depends on the
detailed pattern of the symmetry breaking \cite{Roed}. In recent
years there have been many attempts to use different versions of
the L$\sigma$M to model QCD phase transitions at finite
temperature \cite{BK,Roed,pet}, baryon density
\cite{Scadron,Scavenius}, and for non-equilibrium conditions
\cite{WR,CM,Randrup,{Mocsy:2002hv},bbs}.

Despite many studies the status of the L$\sigma$M remains somewhat
controversial. In most applications of this model only mesonic
degrees of freedom are included explicitly \cite{BK,Randrup,Roed}.
Then the model shows a chiral phase transition, but the
high-temperature phase is very different from what is expected for
QCD, since there are no free quarks but only heavy mesons.
On the other hand, if quark degrees of freedom are included from the 
beginning while fluctuations in the mesonic fields are ignored 
\cite{Scavenius}, the model also predicts a chiral transition at about 
the same temperature. Now, however, the low-temperature phase has the
wrong structure since it is dominated by constituent quarks and,
in the mean field approximation, pions play no role. Physically, of course, 
pions are expected to be the most relevant degrees of freedom at low 
temperatures. There have been several attempts to improve matters by 
including both constituent quarks and pions (see e.g. Ref. \cite{Scadron}), 
but the calculations have been limited to the lowest order loop expansion.
A more satisfactory approach has been pursued recently, namely the 
(approximate) solution of the renormalization group flow equations 
\cite{bjw,wam} which include effects due to field fluctuations. This
appears to be particularly valuable in the neighborhood of critical points.

In the present paper we study a different, and possibly more transparent, 
approach to including field fluctuations. We deal with the full L$\sigma$M
including both constituent quarks and mesonic excitations and our
goal is to proceed as far as possible without invoking any kind of
mean-field approximation or perturbative expansion. We shall
demonstrate below that one can indeed develop a practical
computational scheme where the field fluctuations are
incorporated in the thermodynamic potential to all orders in a
self-consistent way. This formalism corresponds to summing up the
infinite series of daisy and superdaisy diagrams. Of course, this
is possible only within the Hartree approximation where the
exchange diagrams are disregarded. The resulting physical picture
appears to be close to QCD-based expectations, e.g. Ref.
\cite{karsch}. Namely, the model exhibits a smooth chiral crossover 
transition at temperatures of 150--200 MeV from a low-temperature
phase made of heavy constituent quarks and light pions to a
high-temperature phase composed of light quarks and heavy mesonic
excitations. We regard these results as quite satisfactory for
modeling QCD, even though the model has no confinement. For
effective theories that incorporate both chiral symmetry and
confinement see Ref. \cite{Mocsy:2003qw}.

The paper is organized as follows. In Section II a general
L$\sigma$M is formulated and its partition function is represented
within the path integral formalism. We integrate out the quark
degrees of freedom and approximately reduce the problem to a
purely mesonic theory with a very non-linear effective potential.
This potential is then linearized around the correct ground state
including equilibrium field fluctuations to all orders. Finally,
the mesonic contribution to the thermodynamic potential is
calculated in closed form. In Section III we describe a general
method for evaluating averages of complicated functions over the
fluctuations of the fields. This permits easy consistency checks
and allows the equations to be put into a simple form. Results
obtained by including or excluding zero-point contributions, in
addition to the thermal fluctuations, are presented in Section IV.
Here we also assess simplified approximations to the complete
results. Our conclusions are presented in Section V. The
relationship of the present approach to earlier work
\cite{greg3,greg4} is discussed in Appendix A, and the proof of a
useful identity is given in Appendix B.

\section{Theory}
\subsection{L$\sigma$M Lagrangian and Partition Function}

In this paper we employ a standard version of the L$\sigma$M model
with $SU(2)_L\times SU(2)_R$ symmetry. The corresponding
Lagrangian for quarks interacting with the $\sigma$ and $\bm{\pi}$
meson fields is written as
\begin{equation}
{\cal L}={\cal L}_{\bar{q}q}+{\cal L}_{Km}-U(\sigma,\bm{\pi})\;,
\label{lag}
\end{equation}
where the quark Lagrangian is
\begin{equation}
{\cal L}_{\bar{q}q}=
\bar{q}\left[i\gamma_\mu\partial^\mu-
g(\sigma+i\gamma_5\bm{\tau}\cdot\bm{\pi})\right]q\;,
\end{equation}
the meson kinetic energy is
\begin{equation}
{\cal L}_{Km}=\thalf\left(\partial_\mu\sigma\partial^\mu\sigma
+\partial_\mu\bm{\pi}\cdot\partial^\mu\bm{\pi}\right)\;,
\end{equation}
and the meson potential is given by
\begin{equation}
U(\sigma,\bm{\pi})=\tquar\lambda\left(\sigma^2+\bm{\pi}^2-\zeta
\right)^2-H\sigma\, .\label{U}
\end{equation}
Here we have included an explicit chiral symmetry breaking term of
the conventional type, $H\sigma$. As is well known, the choice of
the symmetry breaking term is not unique. The parameters of the
model, $g$, $\lambda$, $\zeta$ and $H$, will be specified later.
The partition function of the grand canonical ensemble can be
written as a functional integral over the quark and meson fields
\cite{kap}
\begin{equation}
Z={\rm Tr}\exp\left\{-\beta(H-\mu\hat{N})\right\}
=\int{\cal D}\bar{q}\,{\cal D}q{\cal D}\sigma{\cal D}\bm{\pi}
\exp\left\{\int\limits_0^\beta d\tau\int\limits_Vd^3x({\cal
L}+\mu\bar{q}\gamma^0q)\right\}\;, \label{part}
\end{equation}
where $\hat{N}$ is the quark number operator, $\beta=T^{-1}$ is the inverse
temperature, $\mu$ is the quark chemical potential, $V$ is the volume of the
system and $\tau=it$ denotes the imaginary time.

\subsection{Integrating Out the Quarks}

First we integrate out the quark degrees of freedom. The relevant
part of the partition function is
\begin{equation}
Z_{\bar{q}q}=\int{\cal D}\bar{q}\,{\cal D}q
\exp\left\{\int\limits_0^\beta d\tau\int\limits_Vd^3x
\bar{q}\hat{D}q\right\}\;,
\end{equation}
where
\begin{equation}
\hat{D}=-\gamma^0\frac{\partial}{\partial\tau}
+i\bm{\gamma}\cdot\pmb{\nabla}
-g(\sigma+i\gamma_5\bm{\tau}\cdot\bm{\pi})
+\mu\gamma^0\;.
\end{equation}
Formally one can integrate out the quark fields with the result
\begin{equation}
\ln Z_{\bar{q}q}=\int\limits_0^\beta d\tau
\int\limits_Vd^3x\ln\det\hat{D}\;.
\end{equation}
The analysis of such an expression has been discussed in detail by
Fraser \cite{fraser}. Accordingly, one proceeds by moving the
operators to the left and the fields which depend on $\tau$ and
$\bm{x}$ to the right. This generates a series of commutators
which involve the derivatives of the meson fields. Since our
interest is in low energy properties we will discard these
commutator terms, tacitly assuming that the meson mode amplitudes
vary slowly in position and time. Then, evaluating the operator
$\hat{D}$ in a frequency-three-momentum representation we get
\begin{equation}
\ln\det\hat{D}=\frac{\nu_q}{2\beta V}\sum_{{\bf p}n}\left\{
\ln\left[\beta^2(\omega_n^2+(E-\mu)^2\right]+
\ln\left[\beta^2(\omega_n^2+(E+\mu)^2\right]\right\}\;,
\end{equation}
where the Matsubara frequency $\omega_n=(2n+1)\pi T$, the quark
degeneracy $\nu_q$ is 12 for the two flavors employed here, and
$E^2=p^2+m^2$. The quark effective mass is given by
\begin{equation}
m^2=g^2(\sigma^2+\bm{\pi}^2)\;. \label{qm}
\end{equation}
After performing the summation over $n$ we get
$\ln Z_{\bar{q}q}=-\int_0^\beta d\tau\int_Vd^3x\,\Omega_{\bar{q}q}(m)$,
where the quark-antiquark thermodynamic potential density is expressed
as
\begin{equation}
\Omega_{\bar{q}q}(m) =-\frac{\nu_qT}{2\pi^2}\int dp p^2
\left\{\beta E+\ln\left[1+e^{-\beta(E-\mu)}
\right]+\ln\left[1+e^{-\beta(E+\mu)}\right]\right\}\,.\label{omnonp}
\end{equation}
Note that this differs from the standard result in that the mass
depends on the meson fields according to Eq. (\ref{qm}). Thus
the partition function (\ref{part}) can be written as
\begin{equation} \label{part1}
Z=\int{\cal D}\sigma{\cal D}\bm{\pi}
\exp\left\{\int\limits_0^\beta d\tau\int\limits_Vd^3x {\cal L}_m\right\}\;,
\end{equation}
where
\begin{equation} \label{mlag}
{\cal L}_m={\cal L}_{Km}-\tilde{U}(\sigma,\bm{\pi})
\end{equation}
is the effective meson Lagrangian. Here the effective potential,
including the contribution of quarks and antiquarks with effective
mass $m$, is
\begin{equation}
\tilde{U}(\sigma,\bm{\pi})= U(\sigma,\bm{\pi})+
\Omega_{\bar{q}q}(m)\;. \label{toteff}
\end{equation}
The Lagrangian (\ref{mlag}) now contains only meson fields and it
constitutes an effective meson theory with the very nonlinear
interaction potential (\ref{toteff}).

\subsection{Linearization of the Mesonic Action}

Let us consider the equations of motion for meson fields which
follow from the effective meson Lagrangian (\ref{mlag}):
\begin{eqnarray}
\partial^\mu\partial_\mu\sigma+\frac{\partial \tilde{U}}{\partial
\sigma}=0\;,
\label{emsi}\\
\partial^\mu\partial_\mu\vec{\pi}+\frac{\partial \tilde{U}}{\partial
\vec{\pi}}=0\;. \label{empi}
\end{eqnarray}
We average these equations over the meson field fluctuations.
Since the $\sigma$ field is expected to develop a nonvanishing
expectation value $v$, we decompose it as $\sigma=v+\Delta$ where
$\Delta$ is the fluctuating part. By definition the average of
$\Delta$ is zero and this is also true for any odd power of
$\Delta$, i.e. $\langle\Delta^n\rangle=0$ for odd $n$. A similar
remark applies to the pion field. Here and below angle brackets indicate
averaging over the field fluctuations. A practical scheme for
evaluating such averages is discussed in the following section.

As the thermal average of an odd number of fluctuating fields is
zero, only the terms in $\tilde{U}$ with odd powers of $\Delta$
will contribute to Eq. (\ref{emsi}), yielding
\begin{equation}
\left\langle\frac{\partial\tilde{U}(v+\Delta,\bm{\pi})}
{\partial\Delta}\right\rangle=0\;. \label{foreom}
\end{equation}
Since the pion field always occurs as $\bm{\pi}^2$, a single
derivative with respect to a component $\pi_i$ will always yield
an odd number of fluctuating fields and thus the thermal average
automatically vanishes. Notice that (\ref{foreom}) includes all
powers of the fluctuating fields, so the equation of motion
contains fluctuations of the fields to all orders. This makes our
approach different from numerous previous attempts to include
fluctuations where an expansion is made around the mean field
ground state. The second derivatives of the effective potential
pick out the even powers of the fluctuating fields in $\tilde{U}$
and these we identify with the meson masses:
\begin{equation}
m_\sigma^2=\left\langle\frac{\partial^2
\tilde{U}(v+\Delta,\bm{\pi})}{\partial\Delta^2}
\right\rangle\quad;\quad m_\pi^2=\left\langle\frac{\partial^2
\tilde{U}(v+\Delta,\bm{\pi})}{\partial\pi_i^2}\right\rangle\;.
\label{mm}
\end{equation}
We now linearize the complicated effective mesonic potential by
setting
\begin{equation}
\tilde{U}(v+\Delta,\bm{\pi})\rightarrow
\left\langle\tilde{U}(v+\Delta,\bm{\pi})\right\rangle
+\thalf m_\sigma^2(\Delta^2-\langle\Delta^2\rangle)
+\thalf m_\pi^2(\bm{\pi}^2-\langle\bm{\pi}^2\rangle)\;. \label{linac}
\end{equation}
Obviously, this becomes an identity if we average over the field
fluctuations on both sides. The terms on the right containing the
average quantities should be included directly in the total
thermodynamic potential density. The remaining terms containing
$\Delta^2$ and $\bm{\pi}^2$ are combined with the kinetic energy
to give the mesonic partition function:
\begin{equation}
Z_m=\int{\cal D}\Delta{\cal D}\bm{\pi}\exp\left\{
\int\limits_0^\beta d\tau\int\limits_V d^3x\left[{\cal
L}_{Km}-\thalf m_\sigma^2\Delta^2-\thalf
m_\pi^2\bm{\pi}^2\right]\right\}\;. \label{zm}
\end{equation}
Following standard steps \cite{kap} one arrives at the
thermodynamic potential density associated with the meson field
fluctuations:
\begin{eqnarray}
\Omega_m&=&-\frac{\ln Z_m}{\beta
V}\equiv\Omega_\sigma+\Omega_\pi\;,\label{omm}\\
\Omega_\sigma&=&\frac{T}{2\pi^2}\int dp p^2\left\{\thalf\beta
E_\sigma +\ln\left(1-e^{-\beta E_\sigma}\right)\right\}\;,
\label{omsi}\\
\Omega_\pi&=&\frac{3T}{2\pi^2}\int dp p^2
\left\{\thalf\beta E_\pi +\ln\left(1-e^{-\beta
E_\pi}\right)\right\}\;, \label{ompi}
\end{eqnarray}
where $E_\sigma=\sqrt{p^2+m_\sigma^2}$ and
$E_\pi=\sqrt{p^2+m_\pi^2}$. Two consistency relations for the
meson masses follow directly from Eq. (\ref{zm}):
\begin{equation} \label{fluc}
\langle\Delta^2\rangle=2\frac{\partial \Omega_\sigma}{\partial
m_\sigma^2}\quad;\quad \langle\bm{\pi}^2\rangle=2\frac{\partial
\Omega_\pi}{\partial m_\pi^2}\;.
\end{equation}
Finally, we can write the total thermodynamic potential density as
\begin{equation}
\Omega=\left\langle{U}(v+\Delta,\bm{\pi})\right\rangle
+\left\langle\Omega_{\bar{q}q}(m)\right\rangle
-\thalf m_\sigma^2\langle\Delta^2\rangle
-\thalf m_\pi^2\langle\bm{\pi}^2\rangle
+\Omega_m(m_\sigma,m_\pi)\;. \label{omtot}
\end{equation}
The subtraction of the third and fourth terms on the right is
necessary to avoid double counting \cite{lm}.

\section{Evaluation of Field Fluctuations}
\subsection{Averaging Procedure}

Consider now the average of a complicated function ${\cal
O}(v+\Delta,\bm{\pi}^2)$ over the fluctuating fields $\Delta$ and
$\bm{\pi}$. The averaging can be carried out by generalizing the
technique introduced in \cite{glue}. First, we expand the function
in a Taylor series about the point $(v,0)$, then take the average
term by term
\begin{equation}
\left\langle{\cal O}(v+\Delta,\bm{\pi}^2)\right\rangle=
\sum_{k,n}{\cal O}^{(k,n)}(v,0)\left\langle\frac{\Delta^k}{k!}
\frac{\bm{\pi}^{2n}}{n!}\right\rangle\;, \label{tay}
\end{equation}
where
\begin{equation}
{\cal O}^{(k,n)}(a,b)\equiv\frac{\partial^k}{\partial a^k}
\frac{\partial^n}{\partial b^n}{\cal O}(a,b)\;.
\end{equation}
The next step is to reduce a general vertex
$\langle\Delta^k\bm{\pi}^{2n}\rangle$ to powers of
$\langle\Delta^2\rangle$ and $\langle\bm{\pi}^{2}\rangle$. The
necessary counting factors for joining the meson fields at this
vertex in all possible ways \cite{greg3} are:
$\langle\Delta^k\rangle=(k-1)!!\langle\Delta^2\rangle^{k/2}$ for
$k$ even, and zero for $k$ odd. For pions, all species are
equivalent so
$\langle\pi_1^2\rangle=\langle\pi_2^2\rangle=\langle\pi_3^2\rangle
=\oneth\langle\bm{\pi}^2\rangle$ and therefore
$\langle\bm{\pi}^{2n}\rangle=(2n+1)!!
\langle\oneth\bm{\pi}^{2}\rangle^n$. After substituting these
factors in the series (\ref{tay}) we notice that the resulting
averaging is equivalent to performing integrations with the
following Gaussian weighting functions:
\begin{eqnarray}
P_\sigma(z)&=&(2\pi\langle\Delta^2\rangle)^{-1/2}\exp\left(
-\frac{z^2}{2\langle\Delta^2\rangle}\right)\;,\\
P_\pi(y)&=&\sqrt{\frac{2}{\pi}}\left(\frac{3}{\langle\bm{\pi}^2
\rangle}\right)^{\frac{3}{2}}\exp\left(
-\frac{3y^2}{2\langle\bm{\pi}^2\rangle}\right)\;.
\end{eqnarray}
Then, resumming the Taylor series, we obtain
\begin{equation}
\langle{\cal O}(v+\Delta,\bm{\pi}^2)\rangle=
\int\limits_{-\infty}^{\infty}dzP_\sigma(z)\int\limits_{0}^{\infty}
dyy^2P_\pi(y){\cal O}(v+z,y^2)\;. \label{genf}
\end{equation}
This is a general result for any analytic function ${\cal O}$.
Note that $\langle{\cal
O}(v+\Delta,\bm{\pi}^2)\rangle\rightarrow{\cal O}(v,0)$ in the
limit when $\langle\Delta^2\rangle$ and
$\langle\bm{\pi}^2\rangle\rightarrow 0$. The correspondence
between Eq. (\ref{genf}) and the approximate expressions used
in Refs. \cite{greg3,greg4} is discussed in the Appendix. We also need
the derivative of Eq. (\ref{genf}) with respect to some
variable $\alpha$. After two integrations by parts one obtains
\begin{equation}
\frac{\partial}{\partial \alpha}\left\langle{\cal
O}(v+\Delta,\bm{\pi}^2)\right\rangle= \frac{\partial
v}{\partial\alpha}\left\langle\frac{\partial{\cal O}
(v+\Delta,\bm{\pi}^2)}{\partial v}\right\rangle\!+
\frac{1}{2}\frac{\partial \langle\Delta^2\rangle}{\partial\alpha}
\left\langle\frac{\partial^2{\cal
O}(v+\Delta,\bm{\pi}^2)}{\partial \Delta^2}\right\rangle\!+
\frac{1}{2}\frac{\partial
\langle\bm{\pi}^2\rangle}{\partial\alpha}
\left\langle\frac{\partial^2{\cal
O}(v+\Delta,\bm{\pi}^2)}{\partial \pi_i^2}\right\rangle\,
.\label{dgenf}
\end{equation}

Using this equation one can easily check that the derivative of
the total thermodynamic potential density with respect to $v$ is
\begin{equation}
\frac{\partial\Omega}{\partial v}=
\left\langle\frac{\partial\tilde{U}}{\partial v}
\right\rangle+\frac{1}{2}\frac{\partial \langle\Delta^2\rangle}
{\partial v}
\left\{\left\langle\frac{\partial^2{\tilde{U}}}{\partial\Delta^2}
\right\rangle-m_\sigma^2\right\} +\frac{1}{2}\frac{\partial
\langle\bm{\pi}^2\rangle}{\partial v}
\left\{\left\langle\frac{\partial^2{\tilde{U}}}{\partial
\pi_i^2}\right\rangle-m_\pi^2\right\}=0\; . \label{minv}
\end{equation}
This vanishes due to the equation of motion (\ref{foreom}) and the mass
definitions (\ref{mm}). Thus we have the non-trivial result that
$\Omega$ is a minimum with respect to variations in the scalar
condensate $v$, as is required for sensible thermodynamics. This
analysis amounts to the formal justification of the approach of
Refs. \cite{greg3,greg4} where the thermal averages were evaluated
approximately in series form, and therefore thermodynamic consistency
was obtained only approximately.

\subsection{Zero-point Fluctuations}

The general formulae (\ref{omnonp}), (\ref{omsi}) and (\ref{ompi})
include both zero-point and thermal fluctuations. The former
are divergent and thus require a proper renormalization procedure.
The non-trivial issue of renormalization in self-consistent
approximation schemes has been discussed for the L$\sigma$M model
by several authors \cite{ris,vers,van}. We adopt their result
for the regularization in four dimensions, namely
\begin{eqnarray}
2\frac{d\Omega_\sigma^{\rm zpt}}{dm_\sigma^2}&=&\frac{1}{4\pi^2}
\int\limits_0^\infty dp\frac{p^2}{E_\sigma}\rightarrow
\frac{1}{16\pi^2}\left[m_\sigma^2\ln\frac{m_\sigma^2}{\Lambda_m^2}
+\Lambda_m^2 -m_\sigma^2\right]\;,\nonumber\\
2\frac{d\Omega_\pi^{\rm zpt}}{dm_\pi^2}&=&\frac{3}{4\pi^2}
\int\limits_0^\infty dp\frac{p^2}{E_\pi}\rightarrow
\frac{3}{16\pi^2}\left[m_\pi^2\ln\frac{m_\pi^2}{\Lambda_m^2}
+\Lambda_m^2 -m_\pi^2\right]\;.
\end{eqnarray}
Note that the $+\Lambda_m^2$ term may be removed by redefinition
of the constant $\zeta$ in the sigma model potential (\ref{U}).
The zero-point contributions to $\Omega$ then follow by
integration (see Eqs.~(\ref{romsi}) and (\ref{rompi}) below).

For the quark case it is natural to adopt a similar form, setting
\begin{equation}
2\frac{d\Omega_{\bar{q}q}^{\rm zpt}}{dm^2}=-\frac{\nu_q}{2\pi^2}
\int\limits_0^\infty dp\frac{p^2}{E}\rightarrow
-\frac{\nu_q}{8\pi^2}\left[m^2\ln\frac{m^2}{\Lambda_q^2}
+\Lambda_q^2-m^2\right]\;.
\end{equation}
As is reasonable in a phenomenological model we have allowed the
renormalization scale $\Lambda_q$ in the fermion sector to differ
from the scale $\Lambda_m$ in the meson sector to allow for the
different physics in the two sectors. Again, integration gives the
result for $\Omega_{\bar{q}q}$ (see Eq.~(\ref{omnon}) below).

\subsection{Practical Evaluation}

The thermodynamic potential densities including both zero-point and
thermal fluctuations of the sigma and pion fields are, respectively:
\begin{eqnarray}
\Omega_{\sigma}&=&\frac{1}{64\pi^2}\left[m_\sigma^4
\ln\frac{m_\sigma^2}{\Lambda_m^2}
+\thalf(3m_\sigma^2-\Lambda_m^2)(\Lambda_m^2-m_\sigma^2)\right]
+\frac{T}{2\pi^2}\int dp p^2\ln\left(1-e^{-\beta E_\sigma}\right)
\;,\label{romsi}\\ \Omega_{\pi}&=&\frac{3}{64\pi^2}
\left[m_\pi^4\ln\frac{m_\pi^2}{\Lambda_m^2}
+\thalf(3m_\pi^2-\Lambda_m^2)(\Lambda_m^2-m_\pi^2)\right]
+\frac{3T}{2\pi^2}\int dp p^2\ln\left(1-e^{-\beta E_\pi}\right)\;.
\label{rompi}
\end{eqnarray}
The squared fluctuations of the meson fields are found by using
Eq. (\ref{fluc}):
\begin{eqnarray}
\langle\Delta^2\rangle&=&2\frac{\partial\Omega_m} {\partial
m_\sigma^2}=\frac{1}{16\pi^2}\left[m_\sigma^2
\ln\frac{m_\sigma^2}{\Lambda_m^2}+\Lambda_m^2-m_\sigma^2\right]
+\frac{1}{2\pi^2}\int dp \frac{p^2}
{E_\sigma}n_B(E_\sigma)\;,\label{sif}\\
\langle\bm{\pi}^2\rangle&=&2\frac{\partial\Omega_m} {\partial
m_\pi^2}=\frac{3}{16\pi^2}
\left[m_\pi^2\ln\frac{m_\pi^2}{\Lambda_m^2}
+\Lambda_m^2-m_\pi^2\right] +\frac{3}{2\pi^2}\int dp \frac{p^2}
{E_\pi}n_B(E_\pi)\;, \label{pif}
\end{eqnarray}
where the Bose occupation number is $n_B(x)=(\exp{\beta
x}-1)^{-1}$. The thermodynamic potential density associated with
quarks is:
\begin{eqnarray}
\langle\Omega_{\bar{q}q}(m)\rangle
&=&-\frac{\nu_q}{32\pi^2}\left\langle
m^4\ln\frac{m^2}{\Lambda_q^2}
+\thalf(3m^2-\Lambda_q^2)(\Lambda_q^2-m^2)\right\rangle\nonumber\\
&&\qquad-\frac{\nu_qT}{2\pi^2}\int dp p^2
\left\langle\ln\left[1+e^{-\beta(E-\mu)}\right]
+\ln\left[1+e^{-\beta(E+\mu)}\right]\right\rangle\;,\label{omnon}
\end{eqnarray}
where $E=\sqrt{p^2+m^2}$.

Since the total thermodynamic potential (\ref{omtot}) is a minimum
with respect to variations in the scalar condensate $v$, the
thermodynamic quantities can be found in standard fashion. The
pressure $P=-\Omega$ and the energy density is
\begin{eqnarray}
{\cal E}&=&\left\langle U(v+\Delta,\bm{\pi})\right\rangle -\thalf
m_\sigma^2 \langle\Delta^2\rangle-\thalf
m_\pi^2\langle\bm{\pi}^2\rangle
+\frac{1}{64\pi^2}\left[m_\sigma^4\ln\frac{m_\sigma^2}{\Lambda_m^2}
+\thalf(3m_\sigma^2-\Lambda_m^2)(\Lambda_m^2
-m_\sigma^2)\right]\nonumber\\
&&+\frac{3}{64\pi^2}\left[m_\pi^4\ln\frac{m_\pi^2}{\Lambda_m^2}
+\thalf(3m_\pi^2-\Lambda_m^2)(\Lambda_m^2-m_\pi^2)\right]
+\frac{1}{2\pi^2}\int dp p^2 \left[E_\sigma n_B(E_\sigma) +3E_\pi
n_B(E_\pi)\right] \nonumber\\ &&-\frac{\nu_q}{32\pi^2}\left\langle
m^4\ln\frac{m^2}{\Lambda_q^2}
+\thalf(3m^2-\Lambda_q^2)(\Lambda_q^2-m^2)\right\rangle
+\frac{\nu_q}{2\pi^2} \int dp p^2\left\langle
E[n_F(E,\mu)+n_F(E,-\mu)]\right\rangle\;,\label{ef}
\end{eqnarray}
where the Fermi-Dirac occupation number is $n_F(x,y)=[\exp{\beta
(x-y)}+1]^{-1}$. Here we do not explicitly show a constant which
must be subtracted from the pressure and added to the energy
density in order to render $P$ and ${\cal E}$ zero in the vacuum.
The quark number density is
\begin{equation}
n=\frac{\nu_q}{2\pi^2}\int
dpp^2\left\langle n_F(E,\mu)-n_F(E,-\mu)\right\rangle\;.
\end{equation}
The entropy density can then be obtained from the standard
thermodynamic relation ${\cal S}=\beta({\cal E}+P-\mu n)$.

For the equation of motion (\ref{foreom}) we need the first
derivative of $\Omega_{\bar{q}q}$:
\begin{equation}
\left\langle\frac{\partial\Omega_{\bar{q}q}}{\partial\Delta}
\right\rangle = g^2\left\langle\sigma A(m)\right\rangle\;,
\label{omd1}
\end{equation}
where the function $A(m)$ is defined by
\begin{equation} \label{Am}
A(m)=2\frac{\partial\Omega_{\bar{q}q}}{\partial m^2}
=-\frac{\nu_q}{8\pi^2}\left[m^2\ln\frac{m^2}
{\Lambda_q^2}+\Lambda_q^2-m^2\right]+\frac{\nu_q}{2\pi^2}\int
dp\frac{p^2}{E}[n_F(E,\mu)+n_F(E,-\mu)]\;.
\end{equation}
Within our approach the quark condensate can be easily expressed as
\begin{equation}
\langle\bar{q}q\rangle
=\frac{1}{g}\left\langle\frac{\partial\Omega_{\bar{q}q}}
{\partial\sigma}\right\rangle=g\left\langle\sigma
A(m)\right\rangle\;. \label{cond}
\end{equation}
For the meson masses in Eq. (\ref{mm}) we need second derivatives:
\begin{eqnarray}
\left\langle\frac{\partial^2\Omega_{\bar{q}q}}{\partial\Delta^2}
\right\rangle&=& g^2\left\langle A(m)+2g^2\sigma^2
\frac{\partial A(m)}{\partial m^2} \right\rangle=\frac{g^2}
{\langle\Delta^2\rangle}\left\langle\Delta(v+\Delta)A(m)\right\rangle\;,
\label{omd2}\\
\left\langle\frac{\partial^2\Omega_{\bar{q}q}}{\partial\pi_i^2}
\right\rangle&=& g^2\left\langle A(m)+2g^2\pi_i^2
\frac{\partial A(m)}{\partial m^2} \right\rangle=\frac{g^2}{\langle
\pi^2_i\rangle}\left\langle\pi^2_iA(m)\right\rangle\;, \label{omd3}
\end{eqnarray}
where for the latter equalities we have used Eq. (\ref{genf}). Explicitly
\begin{equation}
\frac{\partial A(m)}{\partial m^2}
=-\frac{\nu_q}{8\pi^2}\ln\frac{m^2}{\Lambda_q^2}
-\frac{\nu_q}{4\pi^2}\int dp\frac{1}{E}[n_F(E,\mu)+n_F(E,-\mu)]\;.
\end{equation}
When deriving Eqs. (\ref{Am}), (\ref{omd2}) and (\ref{omd3})
we have used the fact that $\Omega_{\bar{q}q}(m)$ and $A(m)$ are
even functions of $m$. For simplicity we have suppressed the
dependence of these functions upon $T$ and $\mu$.

Finally, we need to average the bare potential
$U(v+\Delta,\bm{\pi})$ and its derivatives. Using the expression
(\ref{genf}), or more elementary means, one obtains
\begin{eqnarray}
\left\langle U(v+\Delta,\bm{\pi})\right\rangle&=&
\left\langle\tquar\lambda\left(\sigma^2
+\bm{\pi}^2-\zeta\right)^2-H\sigma\right\rangle
=\tquar\lambda\{3(v^2+\langle\Delta^2\rangle)^2
+(v^2+\langle\bm{\pi}^2\rangle)^2\nonumber\\
&&+\twoth\langle\bm{\pi}^2\rangle^2
+2\langle\Delta^2\rangle\langle\bm{\pi}^2\rangle
-2\zeta(v^2+\langle\Delta^2\rangle+\langle\bm{\pi}^2\rangle
-\thalf \zeta) -3v^4\}-Hv\,. \label{pote}
\end{eqnarray}
The first derivative of $U$ needed in the equation of motion
(\ref{foreom}) is
\begin{equation}
\left\langle \frac{\partial U(v+\Delta,\bm{\pi})}{\partial\Delta}
\right\rangle= \lambda v(v^2+3\langle\Delta^2\rangle
+\langle\bm{\pi}^2\rangle-\zeta)-H\;. \label{du}
\end{equation}
The second derivatives of $U$ needed for the meson masses in
Eq. (\ref{mm}) are
\begin{eqnarray}
\left\langle\frac{\partial^2U(v+\Delta,\bm{\pi})}{\partial\Delta^2}
\right\rangle&=& \lambda(3v^2+3\langle\Delta^2\rangle+
\langle\bm{\pi}^2\rangle-\zeta)\;,\nonumber\\
\left\langle\frac{\partial^2U(v+\Delta,\bm{\pi})}{\partial\pi_i^2}
\right\rangle&=& \lambda(v^2+\langle\Delta^2\rangle+
\fiveth\langle\bm{\pi}^2\rangle-\zeta) \;. \label{ddu}
\end{eqnarray}

The equation of motion (\ref{foreom}) involves the sum of Eqs.
(\ref{omd1}) and (\ref{du}). It is interesting to note that the
quark contribution of Eq. (\ref{omd1}) vanishes if $v=0$. This is
because $\langle\sigma A(m)\rangle$ becomes $\langle\Delta
A(g\sqrt{\Delta^2+\bm{\pi}^2})\rangle$ which is an odd function of
$\Delta$. In the absence of  explicit symmetry breaking, $H=0$,
Eq. (\ref{du}) permits a solution $v=0$ which is the
true solution above the critical point.

The equation of motion and the equations for the meson masses have
to be solved self-consistently. The integrals in Eq. (\ref{genf}) were
evaluated numerically using 32-point Gaussian integration
\cite{as}. The necessary thermodynamic integrals were obtained by
using the numerical approximation scheme of Ref. \cite{johns}.

Assembling the contributions to the meson masses in
Eq.~(\ref{mm}) and using the equation of motion (\ref{foreom}) for
the case $v\neq 0$, we find
\begin{eqnarray}
m_\sigma^2&=&2\lambda v^2+g^2\left\langle\left(
\frac{\Delta}{\langle\Delta^2\rangle}-\frac{1}{v}\right)(v+\Delta)A(m)
\right\rangle+\frac{H}{v}\;,\label{msigf}\\
m_\pi^2&=&2\lambda\left(\langle\pi_i^2\rangle-
\langle\Delta^2\rangle\right)+g^2\left\langle
\left(\frac{\pi_i^2}{\langle\pi_i^2\rangle}-\frac{v+\Delta}{v}
\right)A(m)\right\rangle+\frac{H}{v}\label{mpif}\;.
\end{eqnarray}
It is interesting to consider the high temperature limit when
$\langle\Delta^2\rangle$ and $\langle\pi_i^2\rangle$ become equal.
In this limit the first term on the right in eq.~(\ref{mpif})
vanishes, but so does the second quark term too (see Appendix B).
Therefore, we regain the usual high temperature result,
$m_\pi\simeq H/v$. Since in this region $v$ is very small, the
quark contribution in (\ref{msigf}) also vanishes approximately
and the first term in the equation can be neglected, returning
$m_\sigma\simeq m_\pi$ as expected.

\section{Results}
\subsection{Choice of Parameters}

We need to choose the parameters $\lambda$, $\zeta$, $H$, $g$,
$\Lambda_m$ and $\Lambda_q$ in the case where zero-point
fluctuations are included (labeled ZPT). We also present calculations 
where zero-point fluctuations are excluded (labeled NOZPT) for two 
reasons. Firstly they have often been excluded in the literature and 
secondly it is important to assess their influence in the ZPT case. 
Note that the parameters $\Lambda_m$ and $\Lambda_q$ are irrelevant 
to the NOZPT analysis.

Since chiral symmetry is spontaneously broken in the vacuum, the
axial current requires that $v_{\rm vac}=f_\pi$, where the pion
decay constant is $f_\pi=92.4~$MeV. Goldstone's theorem states
that in the absence of explicit symmetry breaking, $H=0$, the
pion, as a Goldstone boson, should have zero mass in the phase
with spontaneously broken symmetry. Then Eq.~(\ref{mpif}) gives 
in the vacuum
\begin{equation}
2\lambda\left(\langle\pi_i^2\rangle-
\langle\Delta^2\rangle\right)+g^2\left\langle
\left(\frac{\pi_i^2}{\langle\pi_i^2\rangle}-\frac{v+\Delta}{v}
\right)A(m)\right\rangle=0\;. \label{vacgold}
\end{equation}
It is natural to assume that this equation is also valid in the case 
$m_\pi\neq0$, or, stated differently, to take $H=m_\pi^2f_\pi$.
The solution of Eq. (\ref{vacgold}) at $T=0$ is
$\langle\Delta^2\rangle=\langle\pi_i^2\rangle$ since, as shown in
Appendix B, the quark  contribution vanishes in this limit. This
same condition was used by Lenaghan and Rischke \cite{ris} in the
pure meson sector. For the NOZPT case
$\langle\Delta^2\rangle=\langle\pi_i^2\rangle=0$ in the vacuum so
the equation is automatically satisfied. In the ZPT case  
Eq. (\ref{vacgold}) requires $\Lambda_m$ to be intermediate between
the pion and sigma masses. Choosing vacuum masses $m_\sigma=700~$MeV and
$m_\pi=138~$MeV gives $\Lambda_m=453$ MeV, as listed in Table \ref{zzz}; 
values of similar order
have been employed in Refs. \cite{vers,van,chik}. This should not be 
directly viewed as a cut-off parameter. If the $m_\sigma^4$
term in Eq. (\ref{romsi}) or the $m_\sigma^2$ term in Eq. (\ref{sif}) 
is compared with the result of using a four-dimensional cut-off to 
evaluate the integral, the appropriate cut-off parameter is more 
reasonable, namely $\Lambda_me^{3/4}=960$ 
MeV or $\Lambda_me^{1/2}=748$ MeV respectively. 
\begin{table}[t]
\begin{ruledtabular}
\caption{Model Parameters}
\begin{tabular}{|l|r|r|}
parameter&ZPT&NOZPT\\ \hline $\lambda$&7.114 &27.58\\ $\zeta$
(MeV$^2$)&$-8.733\times10^4$ &7.847$\times10^3$\\ $H$
(MeV$^3$)&1.760$\times10^6$ &1.760$\times10^6$\\ $g$&2.844
&3.387\\ $\Lambda_m$ (MeV)&453.4 &---\\ $\Lambda_q$ (MeV)&951.8
&---\\
\end{tabular}
\label{zzz}
\end{ruledtabular}
\end{table}
The value of the coupling constant $g$ is fixed by the requirement
that the vacuum quark mass be approximately a third of the nucleon
mass. We can define the mass by $M_n=\{\langle
m^n\rangle\}^{1/n}$. It is not {\it a priori} obvious which power
of the mass should be averaged. We examine the two lowest moments
with $n=1$ and $n=2$ which are
\begin{eqnarray}
M_1&=&\langle m\rangle =g\left\langle\sqrt{\sigma^2
+\bm{\pi}^2}\right\rangle\;, \label{m1}\\ M_2&=&\langle
m^2\rangle^{\frac{1}{2}} =g(v^2+\langle\Delta^2\rangle
+\langle\bm{\pi}^2\rangle)^{\frac{1}{2}}\;. \label{m2}
\end{eqnarray}
As we shall see below, the two definitions give masses which are
closely similar, so it is of little consequence which choice is
made. We actually used the second definition, setting $M_2=939/3~$
MeV. To determine $\lambda$ and $\Lambda_q$ in the ZPT case we
require the vacuum $m_\sigma$ to be 700 MeV in Eq. (\ref{msigf}) and
constrain the vacuum quark condensate in Eq. (\ref{cond}).
For the latter we choose the value $\langle\bar{q}q\rangle_{\rm
vac}\equiv\langle\bar{u}u+\bar{d}d\rangle= -2\times(225\ {\rm
MeV})^3$ \cite{rein,tmeiss}. In the NOZPT case the vacuum quark
condensate is zero and only the $\sigma$ mass equation is
required. The value of the scale $\Lambda_q$ in Table \ref{zzz} is 
about twice that in the meson sector, as seems intuitively reasonable.
We remark that if $\Lambda_q$ is chosen to be equal to
$\Lambda_m$, and the quark condensate is predicted, rather than
fitted, it turns out to be more than an order of magnitude smaller
than the physical value used above. Furthermore, even when these
renormalization scales differ, it is not possible to carry out the
fit outlined above with $m_\sigma=600~$MeV. This is why we have chosen
a somewhat larger value of 700 MeV for this poorly known mass. Thus,
the vacuum value of the quark condensate is a strong constraint.
It is also responsible for the quite different values of
$\zeta$ listed in columns two and three of Table \ref{zzz}. These were
obtained from the equation of motion (\ref{foreom}). 

Finally, we mention that we have examined results with a number of
different parameter sets and qualitatively they are all very
similar. Therefore the parameters of Table \ref{zzz} will provide a
representative set of results for this model.

We should further point out that while we require Goldstone's
theorem to be satisfied at $T=0$, this does not guarantee that it
will be fulfilled in general. This is because for $T,\mu\neq 0$
the r.h.s. of Eq. (\ref{mpif}) does not vanish even when the
explicit symmetry-breaking term $H$ is set to zero. The
nonvanishing terms can be cancelled by including an additional
(exchange) diagram \cite{Mocsy:2002hv,kapo} which is missing in
our treatment. However, we focus on the more realistic
case where the pion has its physical vacuum mass, $m_\pi=138~$
MeV. Since we find that the contribution of these nonvanishing
terms is relatively small, we shall keep the form (\ref{mpif}) for
the pion mass so as to preserve the structure of the theory.

\subsection{Full Model}

\begin{figure}[t]
 \includegraphics[width=11truecm]{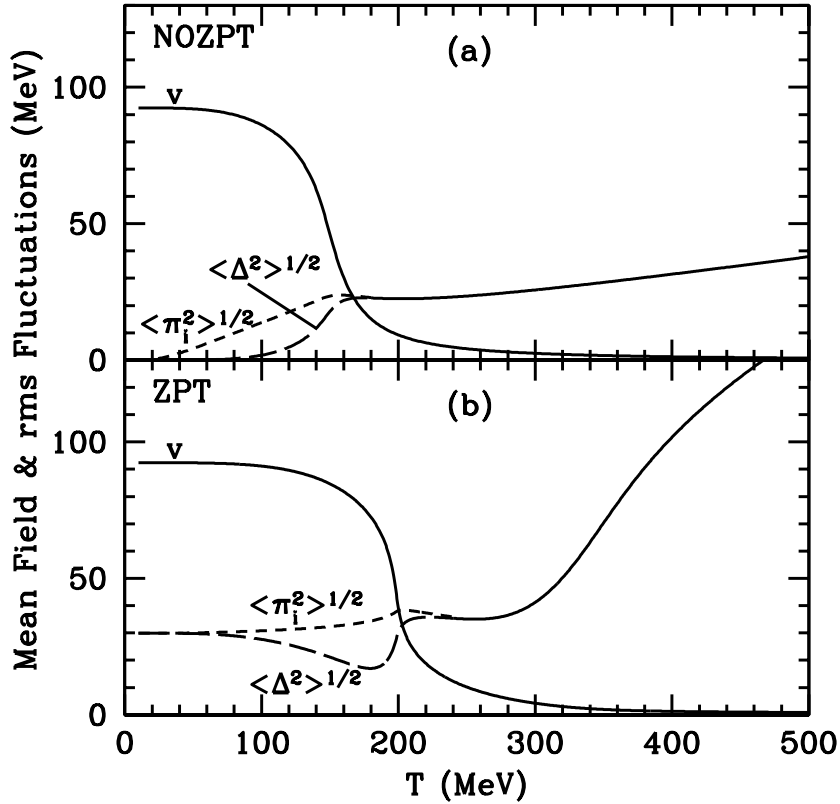}
\caption{The $\sigma$ mean field, $v$, and the root mean square
fluctuations as a function of temperature (a) without and (b) with
zero-point fluctuations.} \label{fone}
\end{figure}

In this paper we consider the case where the quark chemical potential,
$\mu$, is zero, corresponding to a net quark density, $n$, of zero. The
behavior of the sigma mean field $v$ is shown for the NOZPT case
in Fig. \ref{fone}(a) and for the ZPT case in Fig.
\ref{fone}(b). The value of $v$ decreases smoothly to zero as the
temperature increases so that chiral symmetry is approximately restored 
at high temperatures. Thus, there is no sharp phase transition,
but rather a crossover. The crossover temperature, determined from
the maximum value of $|\partial v/\partial T|$, is 150 MeV and
198 MeV in the NOZPT and ZPT cases, respectively.  We remark 
that if quark contributions are omitted this temperature is about 70 MeV
higher in accordance with Ref. \cite{ris}, and in that case $v$
falls off less rapidly with temperature. 
Lattice calculations suggest a smooth crossover, as we have found, 
and extrapolation to the chiral limit yields a phase transiiton 
with a critical temperature of $173\pm8$ MeV \cite{karsch}. The phase 
transition is thought to be of second order, although first order is a 
possibility \cite{PW}. Our model is unlikely to be accurate 
enough to distinguish the order, but it is of interest to examine the
qualitative predictions as we approach the chiral limit. We do this in 
the simplest fashion by multiplying the physical vacuum pion mass by a
factor $<1$ (the ZPT parameters are refitted with the remaining quantities 
unchanged, as specified in Subsec. IVA). We find that when the pion mass 
reduction factor is $\threequ$ the smooth crossover becomes a first 
order phase transition with a critical temperature of 197 MeV. This 
temperature is little changed by further reduction in the pion mass. For 
example with a reduction factor of $\tquar$ the temperature is 194 MeV.
This appears to be an improvement on the prediction of the L$\sigma$M 
without quarks \cite{dum} since it agrees with the lattice finding of a 
weak dependence of the critical temperature on the quark mass \cite{karsch}. 
In comparison the renormalization group calculations \cite{bjw} show a 
somewhat stronger dependence on the pion mass and further a phase 
transition (second order) is obtained only in the chiral limit. The 
transition temperature of 100--130 MeV found in Ref. \cite{bjw} is 
substantially below the values 
given above, no doubt due to the rather small value of 430 MeV 
obtained for the $\sigma$ mass.

\begin{figure}
\includegraphics[width=11truecm]{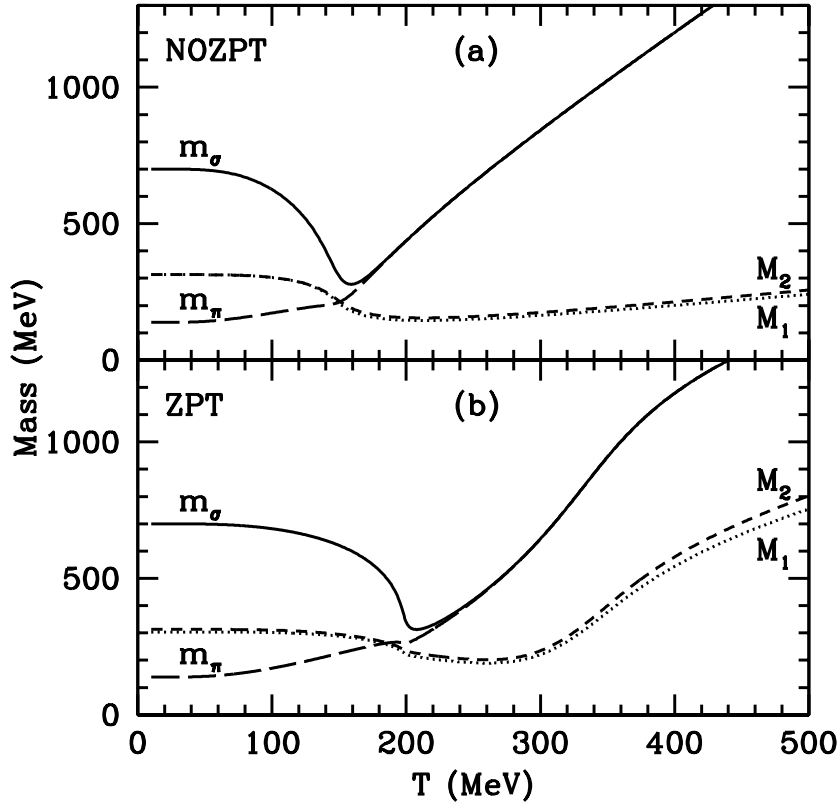}
\caption{The meson and quark masses as a function of temperature
(a) without and (b) with zero-point fluctuations.}
 \label{ftwo}
\end{figure}

Figure \ref{fone} also shows the mean square fluctuations,
$\langle\Delta^2\rangle^{\frac{1}{2}}$ and
$\langle\pi_i^2\rangle^{\frac{1}{2}}$. We required them to be
equal at $T=0$ (see previous subsection), and they are equal again
at high temperatures when $v$ becomes small. As can be seen from
Fig. \ref{fone}(a) the thermal contribution in the intermediate
region is larger for the pion since it has the smaller mass. We
do not see the significant enhancement of the field fluctuations
in the crossover region that would be expected were there to be a 
second order phase transition. The reason is that large thermal
fluctuations are developed only at $m_{\sigma}\ll T$, which never
holds in our calculations: in fact Fig. \ref{ftwo} shows that $m_\sigma$
does not drop below 300 MeV. The shallow dip in
$\langle\Delta^2\rangle^{\frac{1}{2}}$ in the 100--200 MeV temperature 
range which is seen in Fig. \ref{fone}(b) arises from the decrease in 
$m_\sigma$. This causes the zero-point contribution to the
fluctuations in Eq. (\ref{sif}) to decrease, becoming zero at
$m_\sigma=\Lambda_m$. At higher temperatures $m_\sigma$ rises
with increasing $T$, causing the zero-point contribution to
increase strongly and dominate over the thermal contribution for
$T>300~$MeV. This is clearly seen by comparing Figs. \ref{fone}(a) 
and (b).

\begin{figure}
\includegraphics[width=11truecm]{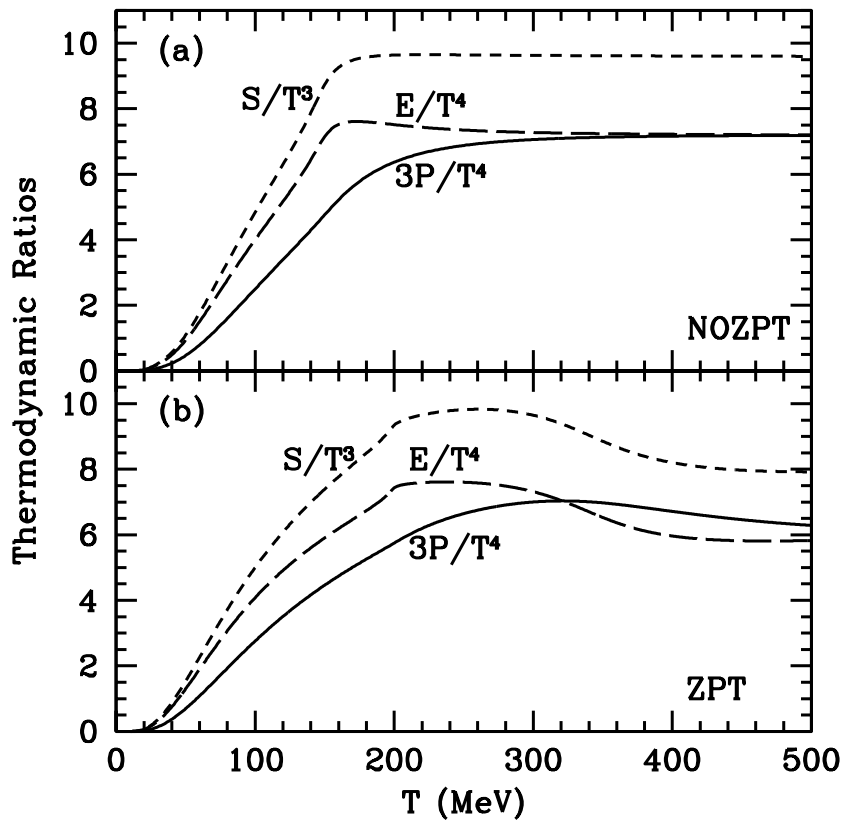}
\caption{The thermodynamic quantities $3P/T^4$, ${\cal E}/T^4$
and ${\cal S}/T^3$ as a function of temperature (a) without and (b) with
zero-point fluctuations.} \label{fthree}
\end{figure}

The behavior of the meson and quark masses is illustrated in
Fig. \ref{ftwo}. For the meson masses the NOZPT and ZPT
calculations give qualitatively similar results. As expected, the
pion mass is an increasing function of $T$ while the sigma mass
develops a minimum in the transition region. Here the pion and
sigma masses become nearly equal signaling the approximate
restoration of chiral symmetry. At high temperatures, $T>250~$MeV,
the meson masses become large and grow according to $\sqrt{H/v}$.
For the quark mass it is seen from Fig. \ref{ftwo} that the
definitions (\ref{m1}) and (\ref{m2}) for $M_1$ (dotted curve) and
$M_2$ (short-dashed curve) give closely similar results. At high
temperatures the mass is small in comparison to the temperature in
the NOZPT case (Fig. \ref{ftwo}(a)), as it is in the renormalization
group approach \cite{bjw}. In contrast in the ZPT case
the increase in the fluctuations seen in Fig. \ref{fone} causes
an increase in the quark mass, such that it is of the same order
as the temperature. Caldas et al. \cite{heron} report a similar
result in their approach to including fluctuations. In this high-$T$
regime the meson masses in Fig. \ref{ftwo}(b) are approximately 
twice the quark mass, as would be expected for loosely bound states 
of quarks and antiquarks.  It should be
emphasized that this behavior is qualitatively different from the
mean-field approximation where the constituent quark mass is
defined to be $g v$, so that it vanishes at high temperatures.

\begin{figure}
\includegraphics[width=10truecm]{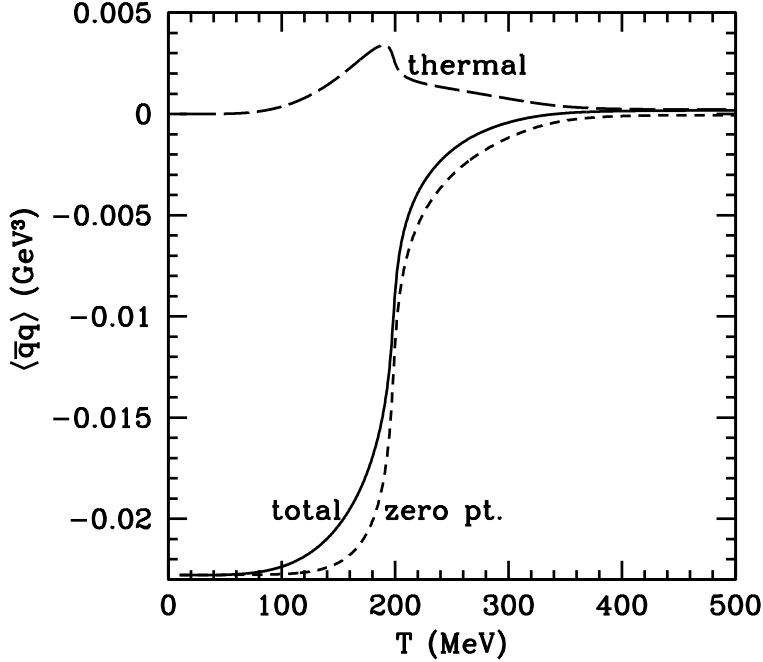}
\caption{Thermal, zero-point and total values of the quark
condensate as a function of temperature in the ZPT case.} \label{ffour}
\end{figure}

The thermodynamic quantities, energy density, pressure and entropy
density, as functions of temperature are presented in Fig.
\ref{fthree}. At high temperatures these are dominated by quarks.
For a massless quark gas of degeneracy 12, $3P/T^4={\cal
E}/T^4=6.91$ and ${\cal S}/T^3=9.21$, and in the NOZPT case in
Fig. \ref{fthree}(a) these values are achieved at high
temperatures. There is only a small contribution from mesons,
since they are heavy, indicating that the system has effectively
become a massless quark gas. The same cannot be said for the ZPT
case in Fig. \ref{fthree}(b), where the asymptotic values are
smaller than the massless limit. This is due the quark mass
remaining comparable to the temperature. 
The quark condensate is displayed in
Fig. \ref{ffour} for the ZPT case since the NOZPT case has only
a small thermal contribution. The condensate is dominated by the
zero-point contribution. Even though one might expect the thermal
part to increase with $T$, the condensate (\ref{cond}) is
approximately $gv\langle A(m)\rangle$, so the decrease in $v$ is
sufficient to bring the thermal contribution to zero at high
temperature. Thus $\langle\bar{q}q\rangle$ starts at $T=0$ with
the chosen empirical value and rapidly becomes negligible when $v$
becomes small. This is the physically expected behavior.

\subsection{Approximations}

The expressions above which involve $\Omega_{\bar{q}q}$ all
require the calculation of two integrals over the Gaussian
weighting functions as well as a momentum integration. Thus it is
worthwhile to examine simpler procedures. The most natural
simplification is to thermally average the quark mass and then
insert it in Eqs. (\ref{omd1}), (\ref{omd2}) and
(\ref{omd3}). We first consider the $M_1$ mass definition in Eq.
(\ref{m1}) which was the approximation adopted in Ref. \cite{greg4}.
Replacing $m$ with $M_1$, the thermodynamic potential density is
approximated
$\langle\Omega_{\bar{q}q}(m)\rangle\rightarrow\Omega_{\bar{q}q}(M_1)$.
In the equation of motion 
\begin{equation}
\left\langle\frac{\partial\Omega_{\bar{q}q}}{\partial\Delta}
\right\rangle=g\left\langle\frac{\sigma}{\sqrt{\sigma^2+\bm{\pi}^2}}
\right\rangle M_1A(M_1)\;,
\end{equation}
and for the meson masses 
\begin{equation}
\left\langle\frac{\partial^2\Omega_{\bar{q}q}}{\partial\Delta^2}
\right\rangle= g\left\langle \frac{\bm{\pi}^2}
{(\sigma^2+\bm{\pi}^2)^{\frac{3}{2}}}\right\rangle
M_1A(M_1)\quad,\quad
\left\langle\frac{\partial^2\Omega_{\bar{q}q}}{\partial\pi_i^2}
\right\rangle= g\left\langle \frac{\sigma^2+\twoth\bm{\pi}^2}
{(\sigma^2+\bm{\pi}^2)^{\frac{3}{2}}}\right\rangle M_1A(M_1)\;.
\end{equation}

\begin{figure}
\includegraphics[width=11truecm]{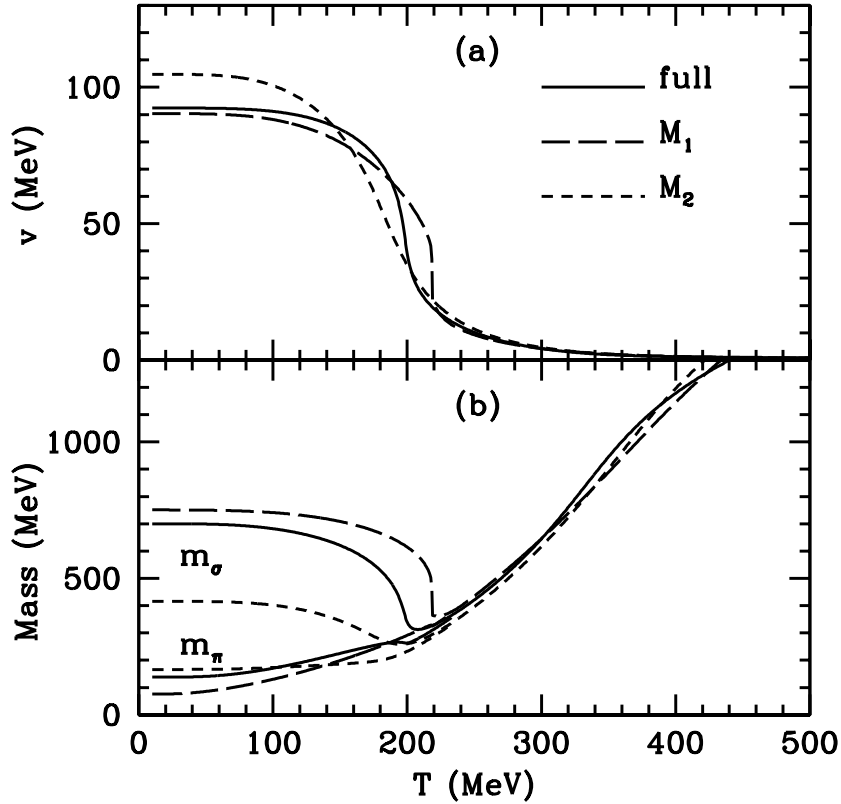}
\caption{Comparison of the full results with the $M_1$ and $M_2$
approximations: (a) $\sigma$ mean field $v$ and (b) meson masses.}
\label{ffive}
\end{figure}
\begin{figure}
\includegraphics[width=11truecm]{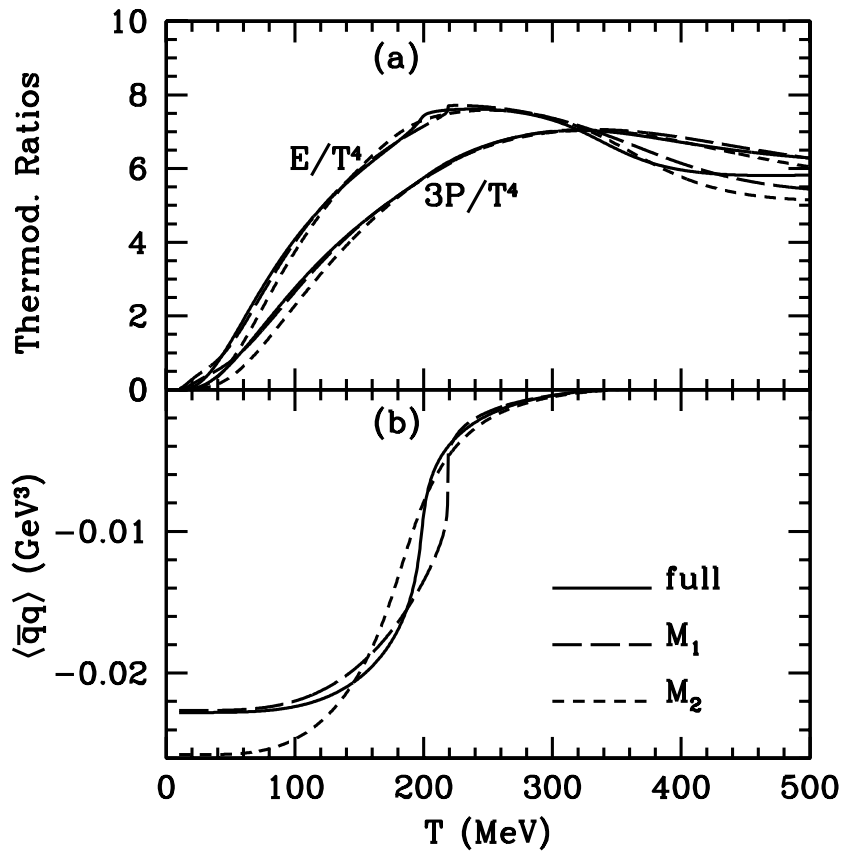}
\caption{Comparison of the full results with the $M_1$ and $M_2$
approximations: (a) thermodynamic quantities $3P/T^4$ and ${\cal
E}/T^4$ and (b) quark condensate.}\label{fsix}
\end{figure}

The $M_2$ case defined in Eq. (\ref{m2}) is even simpler.
Here the thermodynamic potential density is approximated
$\langle\Omega_{\bar{q}q}(m)\rangle\rightarrow\Omega_{\bar{q}q}(M_2)$,
and in the equation of motion
\begin{equation}
\left\langle\frac{\partial\Omega_{\bar{q}q}}{\partial\Delta}
\right\rangle=g^2\langle\sigma\rangle A(M_2)=g^2vA(M_2)\;.
\end{equation}
The second derivatives are $g^2A(M_2)$ for both sigma and pion. In
this case the quark contributions to the meson masses in
(\ref{msigf}) and (\ref{mpif}) vanish.

Note that ambiguity in the evaluation of Eqs. (\ref{omd1}),
(\ref{omd2}) and (\ref{omd3}) when using a thermally averaged quark
mass is resolved by using Eq. (\ref{minv}) to require
that $\partial\Omega/\partial v$ be identically zero. Then the
thermodynamic potential density is a minimum with respect to
variations in the scalar condensate $v$, as it should be.

In Figs. \ref{ffive} and \ref{fsix} we compare the full ZPT
results with those obtained by using the $M_1$ and $M_2$ approximations
for various quantities of interest. The parameters in the left
column of Table \ref{zzz} are used throughout. At temperatures
above 220 MeV both approximations are able to reproduce the exact
results reasonably well with, perhaps, a slight preference for the
$M_1$ case. Below that temperature there are larger deviations
between the exact and the approximate results. The crossover
occurs at a 15 MeV higher temperature in the $M_1$ approximation.
At lower temperatures the $M_1$ approximation gives a reasonable,
although not extremely accurate account of the exact results. The
$M_2$ approximation, on the other hand, shows marked deviations
for the sigma mean field and mass, $v$ and $m_\sigma$, and the
quark condensate. Of course, this can be much improved by
refitting the parameters at $T=0$. However, then one finds that
there are no physical solutions at temperatures just above 300
MeV. We conclude that the $M_1$ approximation is to be preferred
and that it describes the trends of the exact results quite well,
being most accurate at high temperature.

\section{Conclusions}

Our main goal in this paper was to understand the role played by
field fluctuations in effective chiral models. We have shown how
to average field functions of arbitrary complexity over the field
fluctuations. This allowed a set of self-consistent equations for
the average field and the masses to be formulated which led to
consistent thermodynamics. We have applied this approach to the
linear sigma model, including both mesonic and quark degrees of
freedom. The quark degrees of freedom were integrated out and the
effective action was linearized according to Eq.
(\ref{linac}). Thus we described the $\sigma$ and $\pi$ mesons as
quasiparticles and their properties were properly taken into
account in the thermodynamic potential.

We have considered two versions of the model: ZPT where both
zero-point and thermal fluctuations are included, and NOZPT where
only thermal fluctuations are present. The former was able to
describe the quark condensate and the vacuum value was chosen
according to  QCD-based estimates. To accommodate this value a
sigma mass of 700 MeV was employed which, while within the broad
range set by the Particle Data Group \cite{pdg}, is 100 MeV
larger than the traditional value.
We also needed to employ separate renormalization
scales for the mesonic and quark zero-point contributions of 450
MeV and 950 MeV, respectively. For both versions of the model we
required the constituent quark mass in vacuum to be 1/3 of the
free nucleon mass.

Numerical results were presented only for the case of zero net
quark density, i.e. $\mu=0$. The calculations revealed an
interesting and consistent picture. With increasing temperature we
saw a gradual decrease of the condensate and an increase of the
sigma and pion fluctuations. The model gave a crossover type of
chiral transition at a temperature of about 198 MeV (150 MeV) in
the ZPT (NOZPT) case. In the transition region the rms field
fluctuations became comparable in magnitude to the condensate. On the 
other hand we did not observe the strong increase in the sigma field 
fluctuations which would appear in the vicinity of a true critical point.
The restoration of chiral symmetry was seen in the
behavior of the sigma and pion masses, which became degenerate
above the transition region, and in the behavior of the quark
condensate which decreased to zero.

The effective quark mass first showed a modest decrease with
temperature, but above the transition region it started to
increase and this trend was quite strong for the ZPT case. Since
the condensate was already nearly zero, this growth was induced
entirely by the meson field fluctuations. In condensed matter
physics this phenomenon is known as pseudogap formation. At high
temperatures the meson masses increased much more rapidly than the
quark mass, so that they effectively decoupled from the system and we
had a nearly ideal gas of quarks. In the ZPT case the quark
effective mass was comparable to the temperature, however in the
NOZPT case the mass was much smaller so that the thermodynamics
closely resembled a massless quark gas. On the other hand, at low
temperatures quarks were heavy in comparison to the temperature
while pions were relatively light so that we had an ideal gas of
pions with mass close to the physical mass. Certainly, a
qualitatively similar behavior is expected on the basis of QCD.
The transition between these two regimes occurs at a temperature
which is surprisingly close to that found in lattice QCD
simulations \cite{karsch}. In addition we considered simplified
approaches in which the quark mass was treated as a number rather
than a function of the meson fields; these were found to be quite
successful in the high-temperature, chiral-restored regime, but
less accurate at low temperatures.

We also briefly discussed the  dependence of the chiral transition 
on the vacuum pion mass. As the chiral limit was approached by 
reducing the pion mass to a fraction $\leq0.75$ of its true vacuum 
value the crossover transition became a first order phase 
transition. The critical temperature was found to be quite 
insensitive to the value of the pion mass, as in lattice analyses
\cite{karsch}.

In the future it would be interesting to perform calculations for
non-zero quark chemical potential. This would give the possibility
of exploring the phase diagram of the model in the $T$---$\mu$
plane and comparing it with the predictions of other QCD-motivated
models. It would also be interesting to optimize the model in
order to achieve better correspondence with QCD. Some obvious
omissions from the present approach include vector mesons, strange
quarks, the gluon condensate and gluon-like excitations
\cite{glue,impg}. The latter would provide the correct degrees of
freedom at high temperature. Also in the future, more realistic
patterns of symmetry breaking, including for instance the $U(1)_A$ 
anomaly, should be considered.

\section*{Acknowledgements}
This work was supported in part by the Deutsche Forschung
Gemeinschaft (DFG) under grant 436 RUS 113/711/0-1, the US
Department of Energy under grant DE-FG02-87ER40328 and the Russian
Fund of Fundamental Research (RFFR) under grant 03-02-04007. 

\appendix
\section{Connection to Previous Work}

The strategy followed in Refs. \cite{greg3,greg4} was to assume for the
purposes of counting that
\begin{equation}
\langle\Delta^2\rangle=\langle\pi_i^2\rangle=\tquar\langle\Delta^2
+\bm{\pi}^2 \rangle\equiv\tquar\langle\bm{\psi}^2\rangle\;,
\label{alleq}
\end{equation}
although in the numerical evaluation of the final expressions the
actual values of the thermal averages were used. Equation
(\ref{alleq}) is strictly true only in the high temperature limit,
where the sigma and pion masses become degenerate, however at low
temperatures one can also show that the correct expressions are
obtained. Now, writing
$\sigma^2+\bm{\pi}^2=v^2+\bm{\psi}^2+2v\Delta$, the troublesome
cross term, $2v\Delta$, was expanded out. Using (\ref{alleq}) the
expressions needed could be cast in terms of a function
$f(v^2+\bm{\psi}^2)$.

Equation (\ref{alleq}) allows the general expression (\ref{genf})
to be written as a four-dimensional integral for the thermal average
of the function in question:
\begin{equation}
\langle f(v^2+\bm{\psi}^2)\rangle=
\frac{4}{\pi^2\langle\bm{\psi}^2\rangle^2}\int d^4\ell
\exp\left(-\frac{2\ell^2}{\langle\bm{\psi}^2\rangle}\right)
f(v^2+\ell^2)\;.
\end{equation}
Since only the magnitude $\ell^2=\ell_0^2+\ell_1^2+\ell_2^2
+\ell_3^2$ occurs in the integrand, the angular integration may be
carried out giving
\begin{equation}
\langle f(v^2+\bm{\psi}^2)\rangle=
\frac{8}{\langle\bm{\psi}^2\rangle^2}\int
\limits_0^\infty d\ell\ell^3
\exp\left(-\frac{2\ell^2}{\langle\bm{\psi}^2\rangle}\right)
f(v^2+\ell^2)\;.\label{genf4}
\end{equation}

Equation (\ref{genf4}) then allows an easy evaluation of the
principal expressions used in Refs. \cite{greg3,greg4}. Defining a new
integration variable according to $x^2=1+\ell^2/v^2$, then
setting $z^2=2v^2/\langle\bm{\psi}^2\rangle$, and integrating by
parts twice one obtains
\begin{equation}
\left\langle\ln\left(\frac{v^2+\bm{\psi}^2}{v_{\rm vac}^2}\right)
\right\rangle =\ln\left(\frac{v^2}{v_{\rm vac}^2}\right)+1
+(1-z^2)e^{z^2}E_1(z^2)\;,
\end{equation}
as given in \cite{greg3,greg4}, with $v_{\rm vac}$ denoting the
vacuum value of $v$. The exponential integral is defined \cite{as}
by $E_1(y)=\int_1^\infty dt t^{-1}e^{-yt}$. A similar procedure
yields
\begin{equation}
\left\langle\sqrt{v^2+\bm{\psi}^2}\right\rangle=
\frac{3v}{4z}\left[2z+(1-\twoth z^2)\sqrt{\pi}e^{z^2}
{\rm erfc}(z)\right]\;,
\end{equation}
as given in \cite{greg4}, with the complementary error function
defined \cite{as} by erfc$(z)=1-2\pi^{-1/2}\int_0^ze^{-t^2}dt$.

\section{An Identity for the case
$\langle\Delta^2\rangle=\langle\pi_i^2\rangle$}

Here we show that the quark contribution to Eq. (\ref{mpif})
vanishes in the case
$\langle\Delta^2\rangle=\langle\pi_i^2\rangle$. First we prove by
induction that in this case
\begin{equation}
\langle\Delta^{2n+2}\bm{\psi}^{2m}\rangle=(2n+1)\langle\Delta^{2n}
\pi_i^2\bm{\psi}^{2m}\rangle\;, \label{dpeq}
\end{equation}
where $n$ and $m$ are integers. It is simple to verify this
equation in the cases $m=0$ and $m=1$. After integrating by parts
the following relations are obtained
\begin{eqnarray}
\langle \Delta^{2n+2}\bm{\psi}^{2j}\rangle&=&\langle\Delta^2\rangle
\left[(2n+1)\langle \Delta^{2n}\bm{\psi}^{2j}\rangle+2j
\langle \Delta^{2n+2}\bm{\psi}^{2j-2}\rangle\right]\;,\nonumber\\
\langle \Delta^{2n}\pi_i^2\bm{\psi}^{2j}\rangle&=&
\langle\pi_i^2\rangle\left[\langle\Delta^{2n}\bm{\psi}^{2j}\rangle
+2j\langle \Delta^{2n}\pi_i^2\bm{\psi}^{2j-2}\rangle\right]\;.
\end{eqnarray}
Since  $\langle\Delta^2\rangle=\langle\pi_i^2\rangle$ these
relations show that if Eq. (\ref{dpeq}) holds for $m=j-1$
then it holds for $m=j$ and, since it holds for $m=0$ and 1, it
therefore holds in general. The relation can also be proved by
using combinatorial arguments.

Now consider a function $f(m^2)$, where $m^2
=g^2[(v+\Delta)^2+\bm{\pi}^2]$. Expanding $2g^2v\Delta$ in a
Taylor series
\begin{equation}
\langle\Delta f(m^2)\rangle=\sum_{j=0}^\infty\left\langle
\frac{f^{\{2j+1\}}(g^2(v^2+\bm{\psi}^2))}{(2j+1)!}(2g^2v)^{2j+1}
\Delta^{2j+2}\right\rangle\;,
\end{equation}
where we have used the fact that odd powers of $\Delta$ give a
vanishing contribution, and $f^{\{n\}}$ denotes the $n^{\rm th}$
derivative of $f$. Utilizing Eq. (\ref{dpeq})
\begin{eqnarray}
\langle\Delta f(m^2)\rangle&=&2g^2v\sum_{j=0}^\infty\left\langle
\frac{f^{\{2j+1\}}(g^2(v^2+\bm{\psi}^2))}{(2j)!}
\pi_i^2(2g^2v\Delta)^{2j}\right\rangle\nonumber\\
&=&2g^2v\langle\pi_i^2f^{\{1\}}(m^2)\rangle\;,
\end{eqnarray}
again using the fact that odd powers of $\Delta$ give zero
contribution. Thus in the case
$\langle\Delta^2\rangle=\langle\pi_i^2\rangle$ we have the relation
\begin{equation}
\left\langle-\frac{\Delta f(m^2)}{v}+2g^2\pi_i^2\frac{\partial f(m^2)}
{\partial m^2}\right\rangle=
\left\langle\left(\frac{\pi_i^2}{\langle\pi_i^2\rangle}-\frac{v+\Delta}{v}
\right)f(m^2)\right\rangle=0\;, \label{bident}
\end{equation}
where the second equality is obtained using the relation
(\ref{omd3}). Equation (\ref{bident}) was used in the text.


\end{document}